\documentclass[twocolumn,showpacs,preprintnumbers,amsmath,amssymb]{revtex4-1}
\usepackage{amsfonts}
\usepackage{graphicx}
\usepackage{xcolor}

\begin{document}

%\preprint{APS/123-QED}

\title{Stationarity, non-stationarity and early warning signals in economic networks}% Force line breaks with \\

\author{Tiziano Squartini}
\affiliation{Institute for Complex Systems UOS Sapienza, ``Sapienza'' University of Rome, P.le Aldo Moro 5, 00185 Rome (Italy)}
% \altaffiliation[Also at ]{Physics Department, XYZ University.}%%%Lines break automatically or can be forced with \\
\author{Diego Garlaschelli}%
\affiliation{Instituut-Lorentz for Theoretical Physics, Leiden Institute of Physics, University of Leiden, Niels Bohrweg 2, 2333 CA Leiden (The Netherlands)}

\date{\today}

\begin{abstract}
Economic integration, globalization and financial crises represent examples of processes whose understanding requires the analysis of the underlying network structure. Of particular interest is establishing whether a real economic network is in a state of (quasi) stationary equilibrium, i.e. characterized by smooth structural changes rather than abrupt transitions. While in the former case the behaviour of the system can be reasonably  controlled and predicted, in the latter case this is generally impossible. Here we propose a method to assess whether a real economic network is in a quasi-stationary state by checking the consistency of its structural evolution with appropriate quasi-equilibrium maximum-entropy ensembles of graphs. As illustrative examples, we consider the International Trade Network (ITN) and the Dutch Interbank Network (DIN). We find that the ITN is an almost perfect example of quasi-equilibrium network, while the DIN is clearly out-of-equilibrium. In the latter, the entity of the deviation from quasi-stationarity contains precious information that allows us to identify remarkable early-warning signals of the interbank crisis of 2008.
These early-warning signals involve certain dyadic and triadic topological properties, including dangerous `debt loops' with different levels of interbank reciprocity.
\end{abstract}

\pacs{Valid PACS appear here}

\maketitle

\section{Introduction}

Economic and financial systems are strongly interconnected, with several units being linked to each other via different types of interactions. Important examples are provided by trade networks (where economic agents exchange goods or services in return of money) and credit networks (where financial institutions lend and borrow money from each other).

Analysing the intricate structure of economic networks is crucial in order to understand economic dynamics, especially under stress conditions: for instance, the recent global financial crisis has witnessed the role of the increased interconnectedness of the interbank network in the collapse of the system \cite{preferentiallending,may1,subprime,wells,caldarelli_defaults,caldarelli_debtrank,caldarelli_derivatives}. In particular,  while individual banks felt safe minimizing their individual risk by diversifying their portfolios, the simultaneous diversification of all portfolios resulted in an unexpected and uncontrolled level of mutual dependency among all banks, amplifying the effects of individual defaults \cite{may1,subprime}. As another example, understanding the structural organization of international trade networks is crucial in order to measure and characterize economic integration and globalization \cite{giorgio,giorgio2,iman_core}. 

A particularly interesting question is whether the temporal evolution of a real economic network is \emph{quasi-stationary}, i.e. whether the system undergoes smooth structural changes controlled by a few driving parameters. When this is the case, the behaviour of the network is largely controllable and predictable in terms of the dynamics of those parameters. 
On the other hand, the lack of stationarity may result in abrupt transitions and uncontrollable dynamics.

In this extended version of the paper presented at the workshop Complex Networks 2013\footnotetext{Second International Workshop on Complex Networks and their Applications, December 2-5, 2013 Kyoto, Japan. Web page: http://www.complexnetworks.org} (organized within the international conference SITIS 2013\footnotetext{The 9th International Conference on SIGNAL IMAGE TECHNOLOGY {\&} INTERNET BASED SYSTEMS. Web page: www.sitis-conf.org}) and published in the related proceedings \cite{conference_paper}, we address the problem of the (non-)stationarity of real economic networks by studying whether they are found to be typical members of an \textit{evolving quasi-equilibrium ensemble of graphs} with given properties \cite{shannon,jaynes,HL,WF,newman_expo,mylikelihood,mymethod,bargigli-markets}. Roughly speaking, we identify a set of purely topological properties, expected to evolve in time as the natural result of the internal evolution of the network's economic units and we check whether the evolution of the entire network can be simply traced back to the changes in the selected properties. Such properties are treated as \textit{constraints} \cite{jaynes,HL,WF,newman_expo,mylikelihood,mymethod,bargigli-multiplex,caldarelli-boot,caldarelli_credit}, since, in some sense, they are assumed to be the `independent variables' undergoing an autonomous evolution, while the other properties of the network, treated as `dependent variables', are assumed to vary only as a consequence of the former \cite{newman_expo,mylikelihood,mymethod,mywtw}. 

If the observed network properties are systematically found to be in agreement with what expected from the evolving enforced constraints, we can conclude that the real network is quasi-stationary and driven by the dynamics of the constraints.
If the network slightly deviates from the equilibrium expectations, but the deviating patterns are coherent at all times, the network can still be considered consistent with a quasi-stationary one, even if not completely driven by the chosen constraints.
Finally, if the network significantly deviates from the quasi-equilibrium expectation, showing different deviating patterns at different times, then it should be considered a non-stationary one.

We will consider two case studies: the {\it International Trade Network} (ITN), defined as the network of world countries connected by directed import/export relationships (of which we analyse six decades, i.e. 1950, 1960, 1970, 1980, 1990 and 2000) \cite{gleditsch} and the {\it Dutch Interbank Network} (DIN), defined as the network of Dutch banks connected by directed lending/borrowing relationships (of which we consider 44 quarterly snapshots spanning 11 years, i.e. 1998-2008) \cite{mybanks}. 
For simplicity, we will consider both networks in their purely binary and  directed representation, i.e. as graphs where directed links are either present or absent, regardless of their magnitude. We find that, during the considered intervals, the ITN is almost perfectly quasi-stationary, with trade patterns being in systematic agreement with an equilibrium ensemble of graphs specified only by local properties \cite{pre1,mymotifs}. By contrast, the DIN turns out to be strongly non-stationary, displaying different dynamical regimes \cite{mybanks}. As signatures of the major structural changes undergone by the DIN, we find striking early-warning signals of the interbank crisis of 2008.
These early-warning signals are defined in terms of the deviation of certain dyadic and triadic topological properties from their quasi-stationary expectations.
An important role appears to be played by dangerous `debt loops' with different levels of interbank reciprocity.

\section{Quasi-equilibrium graph ensembles}

In this section we introduce the formalism that we use to study the stationarity of real economic networks (but applicable, generally speaking, to any real-world evolving network).

Let us first consider a single (static) snapshot of a real network. 
Such snapshot can be uniquely specified by the \textit{adjacency matrix} $\mathbf{A}$, with entries $a_{ij}=1$ if a link from node $i$ to node $j$ is there, and $a_{ij}=0$ otherwise.
Let us denote the real network by the particular matrix $\mathbf{A}^*$.
Given a set of topological properties that we may choose as constraints (symbolically denoted as $\vec{C}$), it is possible to construct a \textit{statistical ensemble} of graphs, $\mathcal{G}$, such that the expected value $\langle\vec{C}\rangle$ of the constraints over $\mathcal{G}$ is equal to the value $\vec{C}^*$ observed on the real network $\mathbf{A}^*$ \cite{mylikelihood,mymethod}.
The least-biased way to construct this ensemble is that of assigning each graph $\mathbf{A\in\mathcal{G}}$ a probability $P(\mathbf{A})$ such that Shannon's entropy
\begin{equation}
S\equiv -\sum_\mathbf{A\in \mathcal{G}}P(\mathbf{A})\ln P(\mathbf{A})
\end{equation}

\noindent is maximized, under the constraint
\begin{equation}
\langle\vec{C}\rangle=\sum_\mathbf{A\in \mathcal{G}}P(\mathbf{A})\vec{C}(\mathbf{A})=\vec{C}^*
\label{exp}
\end{equation}
where $\vec{C}(\mathbf{A})$ denotes the value of the properties $\vec{C}$ measured on the particular graph $\mathbf{A}$. The solution of the maximization problem is the exponential distribution \cite{shannon}
\begin{equation}
P(\mathbf{A}|\vec{\theta})=\frac{e^{-H(\mathbf{A},\:\vec{\theta})}}{Z(\vec{\theta})},
\end{equation}

\noindent where the so-called \textit{Hamiltonian} $H(\mathbf{A},\:\vec{\theta})\equiv\vec{\theta}\cdot\vec{C}(\mathbf{A})$ is the linear combination of the chosen constraints and the \textit{partition function}, given by $Z(\vec{\theta})\equiv\sum_{\mathbf{A}\in\mathcal{G}}e^{-H(\mathbf{A},\:\vec{\theta})}$, is the normalization constant \cite{jaynes,HL,WF,newman_expo,mymethod}. The parameters $\vec{\theta}$ are the \textit{Lagrange multipliers} that can be set equal to the particular value $\vec{\theta}^*$ such that the expected value of each constraint is equal to the observed one. The value $\vec{\theta}^*$ that realizes eq.(\ref{exp}) can be shown to be also the value that maximizes the log-likelihood  $\ln\mathcal{L}(\vec{\theta})=\ln P(\mathbf{A}^*|\vec{\theta})$ \cite{mylikelihood,mymethod}. 

Once the unknown parameters have been found, it is possibile to evaluate the expected value of any other topological quantity of interest, $X$, as follows 

\begin{equation}
\langle X\rangle^*=\sum_{\mathbf{A}\in\mathcal{G}}X(\mathbf{A})P(\mathbf{A}|\vec{\theta}^*).
\label{eq:X}
\end{equation} 

\noindent If the real network is a typical member of the ensemble, the knowledge of the constraints will be enough to reproduce the original network; otherwise the knowledge of additional properties will be required.
We have recently proposed a completely analytical method allowing one to compare any topological property of the real network with the corresponding expected value over the constructed ensemble, in the fastest possible time \cite{mymethod}.

We now show how it is possible to extend the above ideas to study whether a \textit{dynamically evolving} network is consistent with a quasi-equilibrium ensemble.
Given a temporal sequence $\{\mathbf{A}^*(t)\}_t$ of snapshots of a real network and a set of constraints $\vec{C}$, we have a different observed vector $\vec{C}^*(t)$ for each timestep $t$. Thus, a different maximum-entropy graph ensemble, such that the ensemble average $\langle \vec{C}(t)\rangle$ equals $\vec{C}^*(t)$, can be generated for each timestep $t$. Now, in order to check whether the evolution of the real network is consistent with a quasi-equilibrium process driven by smooth changes in only a small set of its topological properties, $\{\vec{C}(t)\}$ can be taken to be precisely the temporal sequence of desired properties.
Then, by iterating the aforementioned procedure on all snapshots, it can be checked whether the real network's evolution is consistent with that of the quasi-equilibrium ensemble generated by the dynamics of $\vec{C}^*(t)$.

One of the most important examples is when the driving property is the \textit{degree sequence}, i.e. the number of links of each node. If $k_i$ denotes the degree of node $i$, then the vector $\vec{k}$ denotes the degree sequence of the entire network. Specifying the degree sequence as the driving quantity amounts to choose $\vec{C}(t)\equiv \vec{k}(t)$.
Being a completely local property, the degree of a node is the quantity most prone to be interpreted in terms of intrinsic economic features (such as wealth, income, capitalization, etc.) characterizing that node. For instance, the degree of countries in the ITN is strongly and non-linearly correlated with the Gross Domestic Product (GDP) \cite{mywtw}.
In this  interpretation, assessing that an economic network undergoes a quasi-equilibrium evolution driven by the dynamics of its local topological properties (e.g. the degree sequence) allows one to conclude that the network's evolution is driven by the changes of intrinsic node-specific economic variables.

In the Appendix we briefly discuss what are the possible local properties that can be defined in a directed network. This leads us to the introduction of three different ensembles: the \textit{Directed Random Graph Model} (DRG), defined by the \textit{total number of links of the network, $L$,} \cite{mymethod}, the \textit{Directed Configuration Model} (DCM), defined by the \textit{in-degree} and \textit{out-degree sequences}, $k^{out}_i$ and $k^{in}_i$, $\forall\:i$ (i.e. the directed generalization of the concept of degree) \cite{newman_expo,mymethod} and the \textit{Reciprocal Configuration Model} (RCM), defined by the \textit{reciprocated degree}, \textit{non-reciprocated in-degree} and \textit{non-reciprocated out-degree sequences}, $k^{\leftrightarrow}_i$, $k^{\leftarrow}_i$, $k^{\rightarrow}_i$, $\forall\:i$ \cite{mymethod,mymotifs,mygrandcanonical,myreciprocity}.

\section{Dyadic and triadic motifs}

Since they assume that the network arises as a simple combination of purely topological properties, the DRG, the DCM and the RCM are typically treated as \textit{null models}, i.e. simple models expected to fail in reproducing the data, but useful precisely because they can highlight interesting patterns in the real system in terms of deviations from the null hypothesis.
Here, the systematic accordance with a null model throughout the considered time period would indicate a quasi-equilibrium network evolution driven by the constraints defining the null model itself.
In some sense, a good but incomplete accordance could still indicate a quasi-stationary evolution, as long as the deviating patterns were always the same and with the same amplitude. In this case, the dynamics of the network would not be entirely driven by that of the constraints themselves. By contrast, a network out of the quasi-equilibrium dynamics generated by the chosen constraints would display wild and irregular deviations from the null model's expectations. 

\begin{figure}[t!]
\centering
\includegraphics[width=0.25\textwidth]{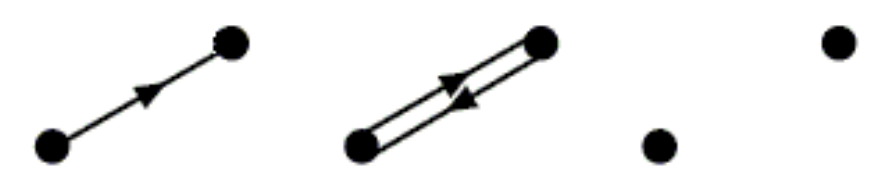}
\caption{The 3 dyadic motifs representing all the possible non-isomorphic topological configurations involving two connected nodes in a directed network.}
\label{fig_dy}
\end{figure}

\begin{figure}[t!]
\centering
\includegraphics[width=0.49\textwidth]{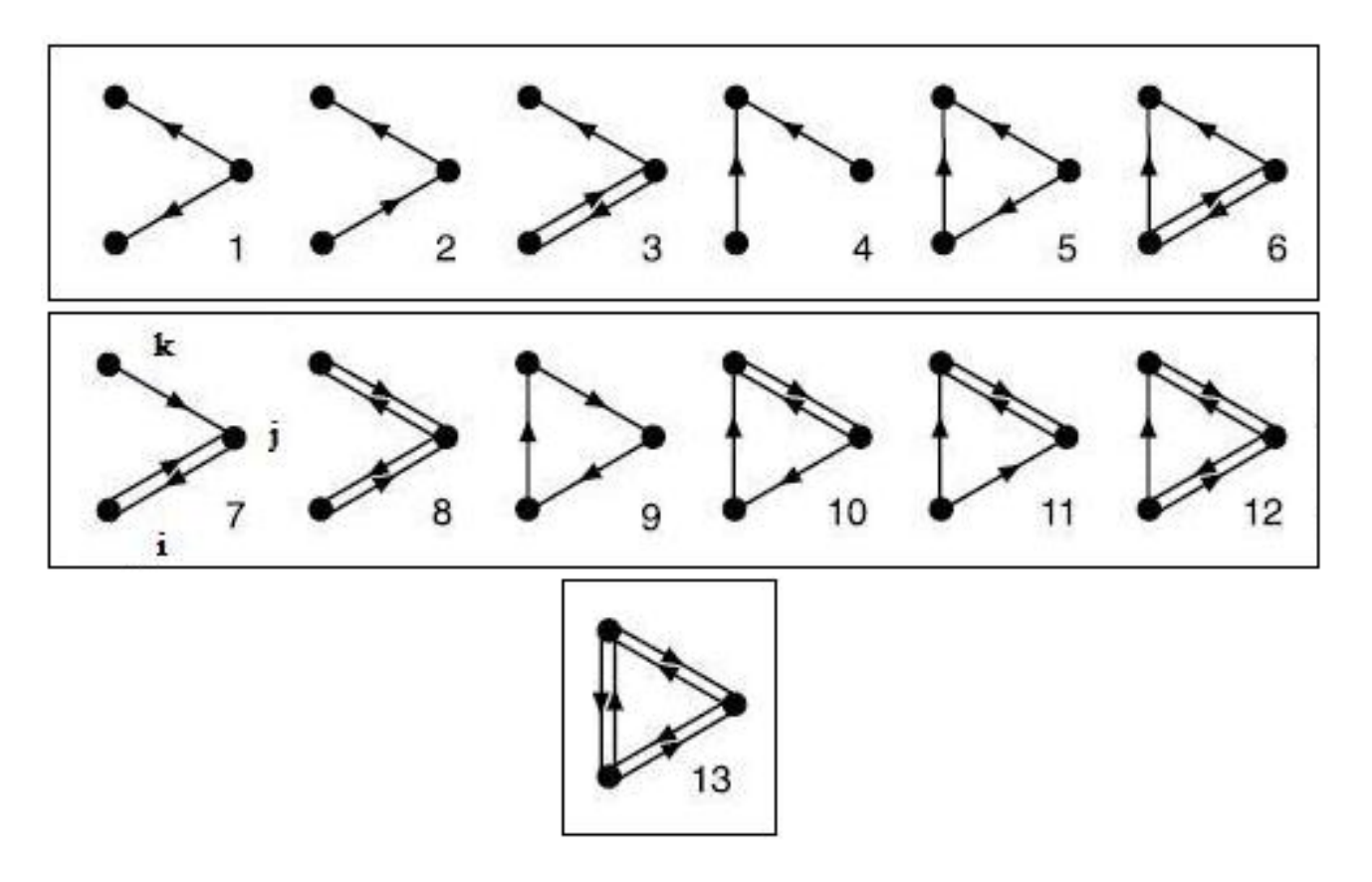}
\caption{The 13 triadic motifs representing all the possible non-isomorphic topological configurations involving three connected nodes in a directed network. \copyright 2014 IEEE. Reprinted, with permission, from Proceedings of the Ninth International Conference on Signal-Image Technology \& Internet-Based Systems (SITIS 2013), pp. 530-537 (edited by IEEE) (2014).}
\label{fig_mot}
\end{figure}

Since the constraints specified in the DRG, the DCM and the RCM are global, node-specific and dyad-specific respectively, the simplest non-trivial (i.e. higher-order) properties to monitor are \textit{dyad-specific} (in the case of the DRG and DCM) and \textit{triad-specific} (in the case of the RCM), i.e. involving, respectively, pairs and triples of nodes. For this reason, in this paper we analyse in detail the so-called \textit{dyadic} and \textit{triadic motifs} \cite{calda_book,foodwebmotifs,motifs,motifs2}.
Dyadic motifs are defined as the 3 non-isomorphic topological configurations involving two connected nodes in directed networks (see Fig. \ref{fig_dy}).
Similarly, triadic motifs are defined as the 13 non-isomorphic topological configurations involving three connected nodes in directed networks (see Fig. \ref{fig_mot}).

The number of occurrences $N_m$ of a particular motif $m$, either dyadic ($m=L^\rightarrow,\:L^\leftrightarrow,\:L^x$) or triadic ($m=1\dots13$), can be written in two equivalent ways. The first one employs products of adjacency matrix elements, $a_{ij}$, and is suitable when using the DRG and the DCM. 
The second one employs the mutually excluding quantities $a_{ij}^{\rightarrow}\equiv a_{ij}(1-a_{ji})$, $a_{ij}^{\leftarrow}\equiv a_{ji}(1-a_{ij})$ and $a_{ij}^{\leftrightarrow}\equiv a_{ij}a_{ji}$ and is particularly useful when using the the RCM. 

For instance, the abundance of the dyadic motif $m=L^{\leftrightarrow}$ can be calculated as

\begin{equation}
N_{L^{\leftrightarrow}}(\mathbf{A})=\sum_{i\neq j}a_{ij}a_{ji}=\sum_{i\neq j}a_{ij}^{\leftrightarrow},
\label{eq:dy_product}
\end{equation}

\noindent Its expected value under the DRG and the DCM reads
\begin{eqnarray}
\langle N_{L^{\leftrightarrow}}\rangle_{DRG}&=&\sum_{i\neq j}p^2=2!\cdot\binom{N}{2}p^2,\nonumber\\
\langle N_{L^{\leftrightarrow}}\rangle_{DCM}&=&\sum_{i\neq j}p_{ij}p_{ji}
\label{eq:dy_product2}
\end{eqnarray}

\noindent while its expected value under the RCM reads

\begin{equation}
\langle N_{L^{\leftrightarrow}}\rangle_{RCM}=\sum_{i\neq j}p_{ij}^{\leftrightarrow}.
\label{eq:dy_product3}
\end{equation}

Equivalently, the abundance of the triadic motif $m=10$ can be calculated as
\begin{eqnarray}
N_{10}(\mathbf{A})&=&\sum_{i\ne j\ne k}(1-a_{ij}) a_{ji} a_{ik} (1-a_{ki}) a_{jk} a_{kj}\nonumber\\
&=&\sum_{i\ne j\ne k}a_{ij}^{\leftarrow} a_{ik}^{\rightarrow} a_{jk}^{\leftrightarrow},
\label{eq:product}
\end{eqnarray}

\noindent Its expected value under the DRG and the DCM is
\begin{eqnarray}
\langle N_{10}\rangle_{DRG}&=&\sum_{i\ne j\ne k}p^4(1-p)^2=3!\cdot\binom{N}{3}p^4(1-p)^2,\nonumber\\
\langle N_{10}\rangle_{DCM}&=&\sum_{i\ne j\ne k}(1-p_{ij}) p_{ji} p_{ik} (1-p_{ki}) p_{jk} p_{kj}
\label{eq:product2}
\end{eqnarray}
\noindent while its expected value under the RCM is
\begin{eqnarray}
\langle N_{10}\rangle_{RCM}=\sum_{i\ne j\ne k}p_{ij}^{\leftarrow} p_{ik}^{\rightarrow} p_{jk}^{\leftrightarrow}.
\label{eq:product3}
\end{eqnarray}

Given a real network $\mathbf{A}^*$, the usual way to compare the  observed and expected abundance of motifs is by means of the so-called \textit{z-scores}, i.e. the standardized quantities 
\begin{equation}
z_{m}\equiv\frac{N_{m}(\mathbf{A}^*)-\langle N_{m}\rangle^*}{\sigma^*[N_{m}]}
\label{eq:z}
\end{equation}
where $\sigma^*[N_{m}]\equiv\sqrt{\langle N_{m}^2\rangle^*-(\langle N_{m}\rangle^*)^2}$ is the standard deviation of $N_{m}$ under the null model.
If the observations were exactly reproduced by the null model, then the $z$-scores would be exactly zero. On the other hand, significantly large positive or negative $z$-scores indicate an over- or under-estimation of the motifs' empirical abundance respectively. The meaning of the $z$-scores is well defined for normally distributed variables (e.g. for dyadic motifs): in this case, the deviations can be nicely quantified in terms of probabilities, as the intervals $z_m=\pm1,\:\pm2,\:\pm3$ select regions enclosing a probability of $68\%$, $95\%$ and $99.7\%$, respectively.
Choosing one of the above values as a threshold allows the identification of significantly deviating patterns. While for non-normally distributed variables (e.g. for triadic motifs) it is impossible to attach probabilities to $z$-scores, large values still highlight the most deviating patterns and their temporal evotion still enables to assess the (non-)stationarity of the network.

Since the values of $z_m$ are sensitive to the number of nodes, when it is necessary to compare the $z$-scores of networks with different size, or of differently sized snapshots of the same network, a size-independent measure is needed. For this reason, it is customary to normalize the $z$-scores by introducing the \textit{significance profile} \cite{motifs,motifs2} defined as
\begin{equation}
SP_{m}\equiv\frac{z_{m}}{\sqrt{\sum_{m=1}^{13}z_{m}^2}}
\label{eq:SP}
\end{equation}

\noindent and measuring the \textit{relative} importance of each motif with respect to the other ones. While the $z$-scores are unbounded quantities, $SP_m$ lies between $-1$ and $+1$.

\section{The International Trade Network}

Equipped with the techniques and formalism described so far, we now start showing the results of the empirical analysis of the first of the two economic networks mentioned in the Introduction, i.e. the ITN.

In the binary, directed representation of the ITN, nodes represent world countries and a directed link from node $i$ to node $j$ represents the existence of an export relation from country $i$ to country $j$.
The initial number (85) of countries roughly doubles during the time period considered (1950-2000), mainly because of many colonies becoming independent and the Soviet Uniot disgregating into many states.
This expansion of the network and the simultaneous globalization process have caused a significant increase in the number of links \cite{mywtw}, as well as considerable variations in the nodes' degrees as well.
This circumstance makes the ITN an ideal example for testing whether an economic network undergoes a quasi-equilibrium evolution driven by the dynamics of the local properties.

\begin{figure}[t!]
\centering
\includegraphics[width=0.393\textwidth]{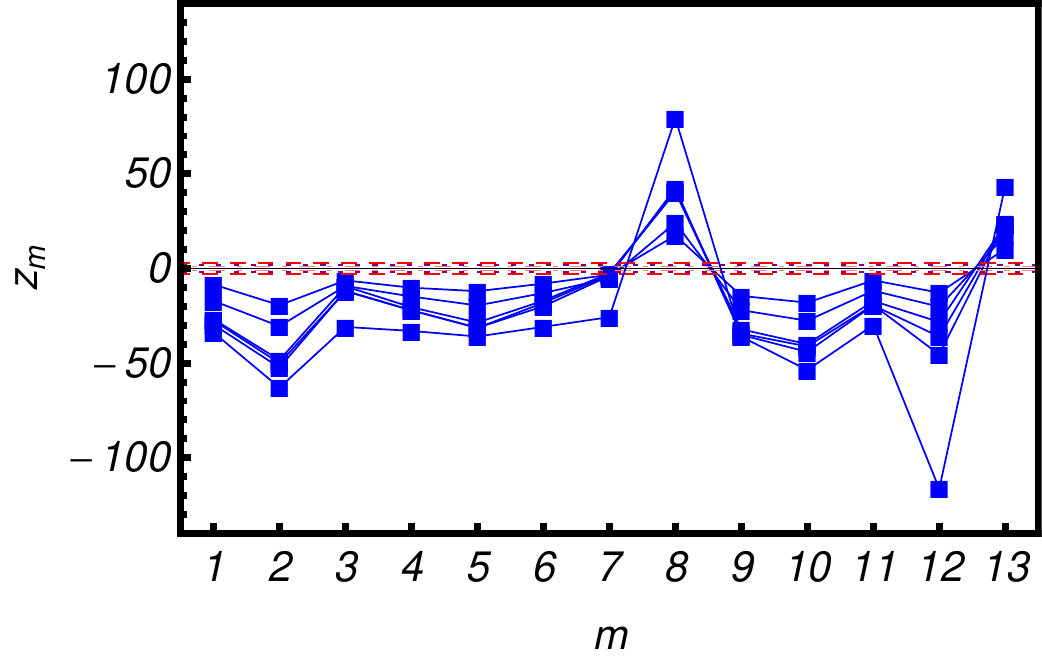}\\
\hspace{3.5mm}\includegraphics[width=0.377\textwidth]{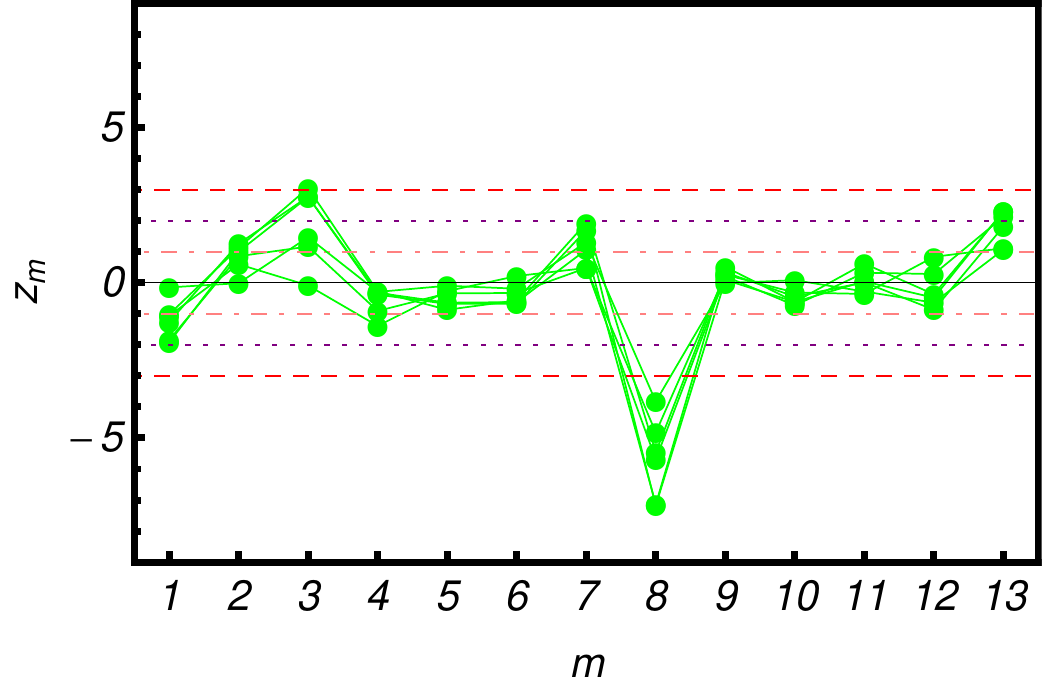}\\
\includegraphics[width=0.395\textwidth]{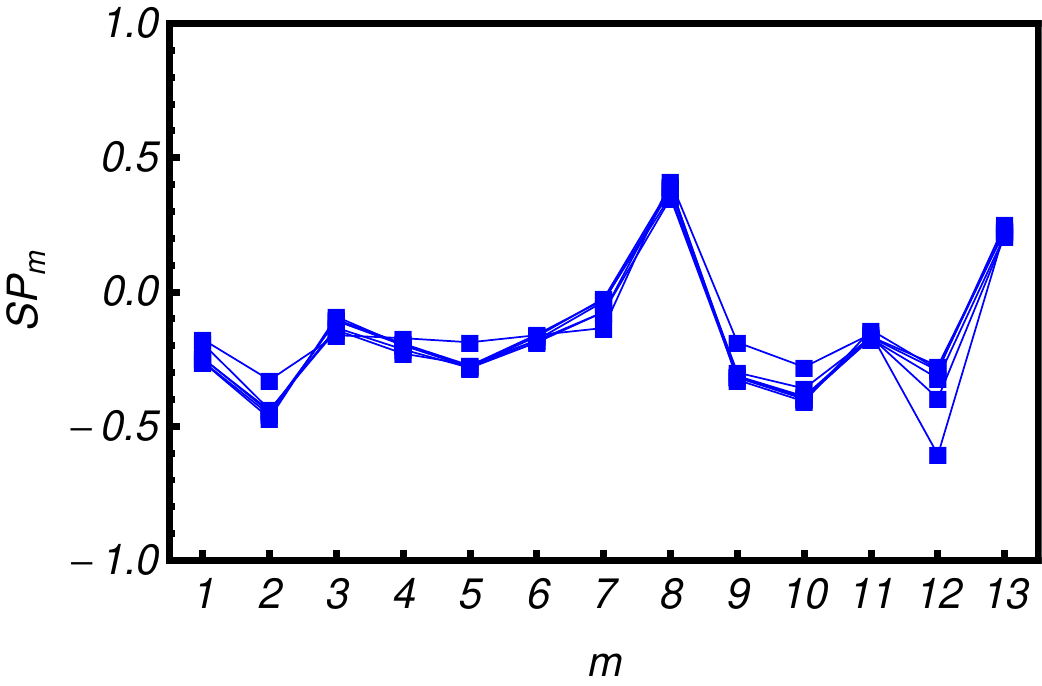}\\
\includegraphics[width=0.395\textwidth]{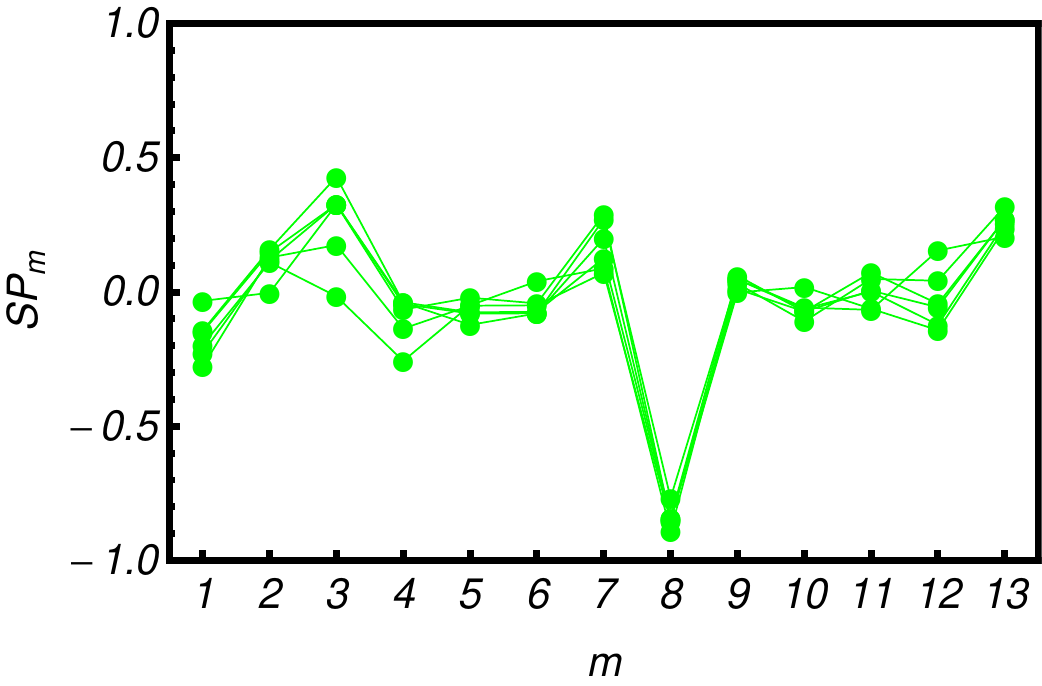}
\caption{$z$-scores (first and second panel) and significance profiles (third and fourth panel) of the 13 triadic, binary, directed motifs for the ITN in the years 1950, 1960, 1970, 1980, 1990 and 2000, under the DCM ($\textcolor{blue}{\blacksquare}$, first and third panel) and the RCM ($\textcolor{green}{\bullet}$, second and fourth panel). The dashed, red lines represent the values $z=\pm3$, the dotted, purple lines the values $z=\pm2$ and the dot-dashed, pink lines the values $z=\pm1$. \copyright 2014 IEEE. Reprinted, with permission, from Proceedings of the Ninth International Conference on Signal-Image Technology \& Internet-Based Systems (SITIS 2013), pp. 530-537 (edited by IEEE) (2014).}
\label{fig_wtw}
\end{figure}

We avoid the use of the DRG as a meaningful null model, on the basis of the following simple considerations. The single parameter of the model, i.e. the probability coefficient $p$, coincides with the link density, which throughout the evolution of the ITN is approximately $p\simeq 1/2$ \cite{pre1}. This means that the DRG would predict a network structure where the presence of each link is determined by simply tossing an almost fair coin. This oversimplified model is completely uninformative, i.e. is almost equivalent to a model where no piece of information is available about the network, and is of course unable to reproduce any property of the real ITN.
Moreover, since the DRG is defined only in terms of the total number of links, or equivalently in terms of the average degree of vertices, interpreting this quantity as a local driving property of nodes amounts to completely neglect the inter-node variability of the economic factors determining the degree of a country.

We therefore focus on the DCM and the RCM. The results of the anaysis of the $z$-scores, as defined in eq.(\ref{eq:z}), are shown in Fig. \ref{fig_wtw}. Under the DCM, the $z$-scores indicate large deviations between observations and expectations, and the agreement worsens as the network evolves. 
These results confirm that, while some higher-order properties of the ITN were previously found to be well-reproduced by constraining the nodes' degrees \cite{pre1}, the triadic patterns are irreducible to the in- and out-degrees themselves \cite{mymotifs}.

By contrast, the agreement improves substantially under the RCM: now, all the $z$-scores (with the only exception of motif 8) lie within the error bars $z_m=\pm 3$. 
This indicates that, once the number of reciprocated and non-reciprocated links of each node are separately controlled for, the triadic structure of the network is almost completely explained.
Moreover, the shape of the profiles is more stable than under the DCM.
All these findings indicate that the RCM should be preferred to the DCM, the reciprocity structure playing a strong role in shaping the topology of the ITN \cite{mywtw,mymotifs,myreciprocity}.

It should in any case be noted that the $z$-scores' profiles display a high degree of stability.
The panels of Fig. \ref{fig_wtw} also show the significance profiles for all 13 motifs, as defined in eq.(\ref{eq:SP}). We find that discounting the effect of the increasing size of the network makes the curves of the 6 different snapshots collapse to a single profile.
This effect is obviously more evident under the DCM, since under the RCM the $z$-scores of the different snapshots were already largely overlapping. 

\begin{figure}[h!]
\centering
\hspace{1mm}\includegraphics[width=0.384\textwidth]{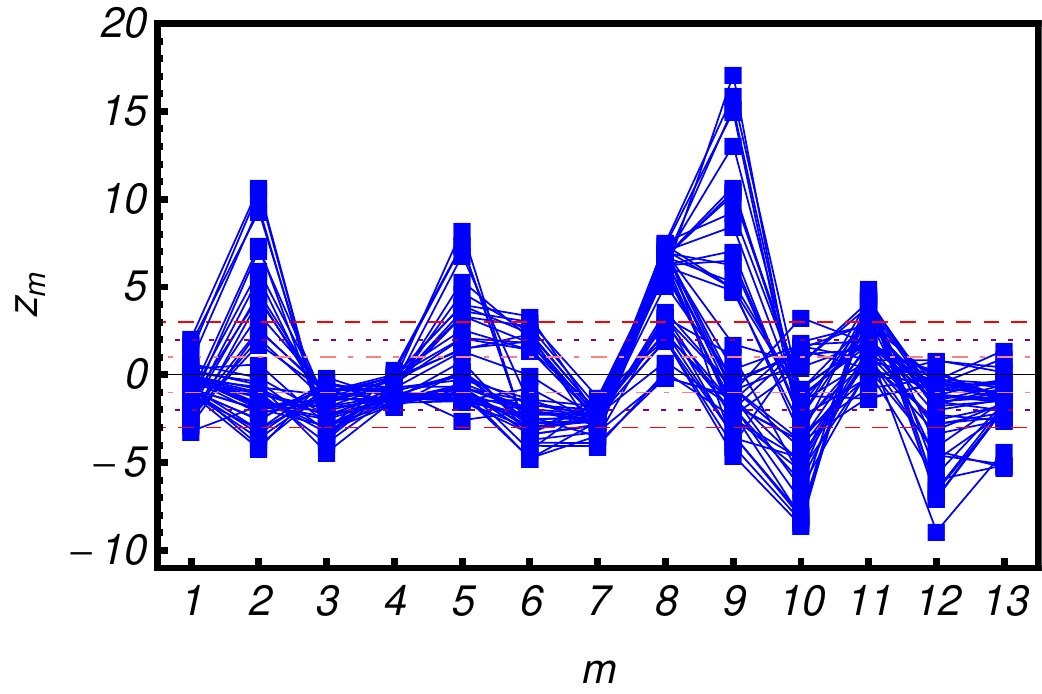}\\
\hspace{2mm}\includegraphics[width=0.377\textwidth]{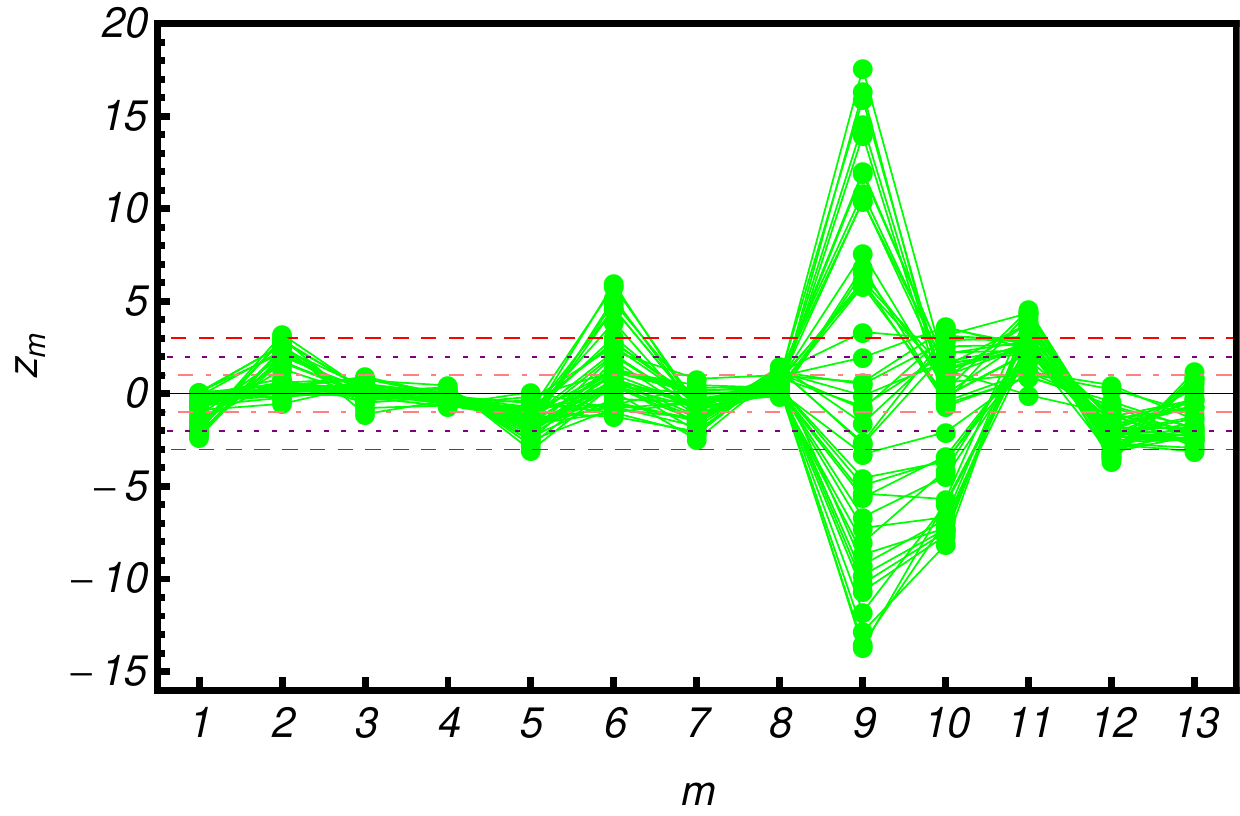}\\
\includegraphics[width=0.389\textwidth]{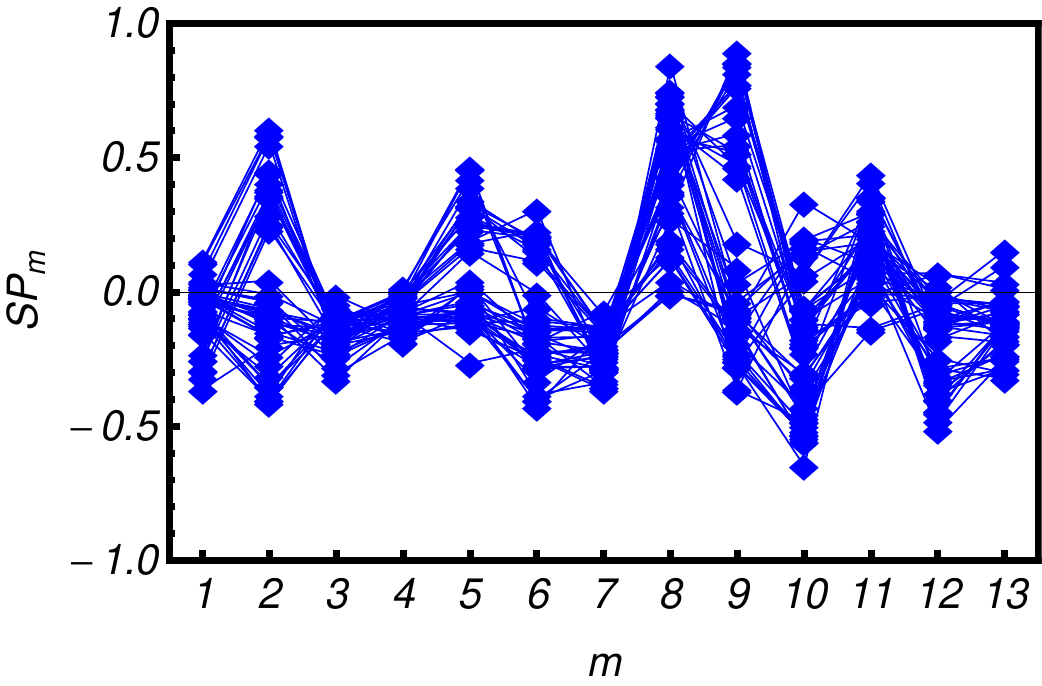}\\
\includegraphics[width=0.389\textwidth]{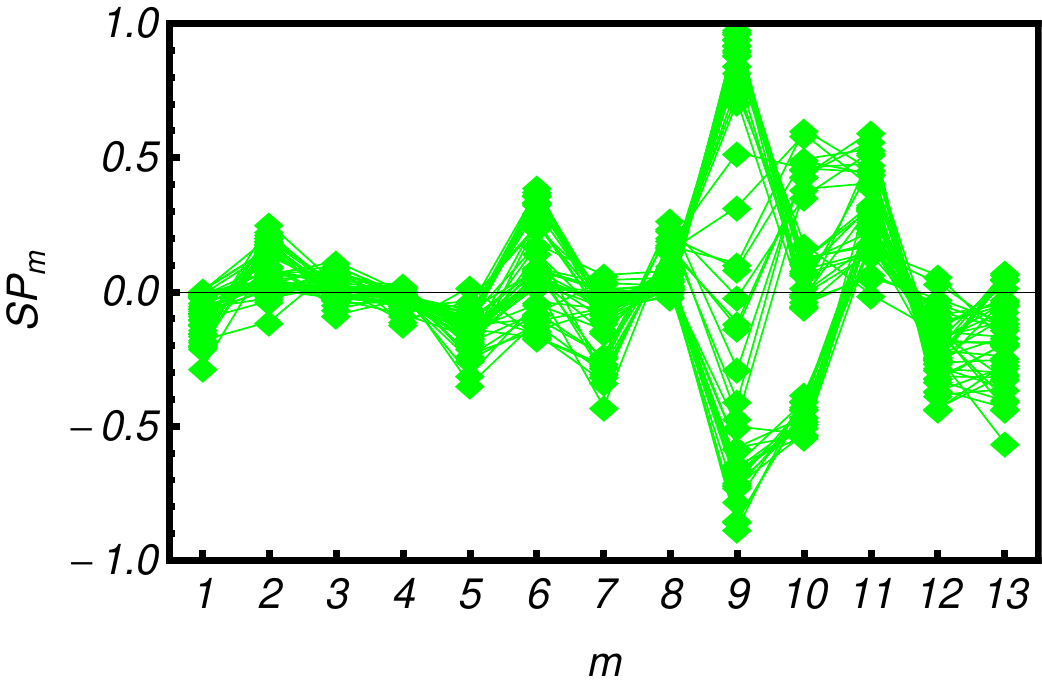}
\caption{$z$-scores (first and second panel) and significance profiles (third and fourth panel) of the 13 triadic, binary, directed motifs for the 44 quarterly snapshots of the DIN between 1998 and 2008, under the DCM ($\textcolor{blue}{\blacksquare}$ and $\textcolor{blue}{\blacklozenge}$, first and third panel) and the RCM ($\textcolor{green}{\bullet}$ and $\textcolor{green}{\blacklozenge}$, second and fourth panel). The dashed, red lines represent the values $z=\pm3$, the dotted, purple lines the values $z=\pm2$ and the dot-dashed, pink lines the values $z=\pm1$. \copyright 2014 IEEE. Reprinted, with permission, from Proceedings of the Ninth International Conference on Signal-Image Technology \& Internet-Based Systems (SITIS 2013), pp. 530-537 (edited by IEEE) (2014).}
\label{fig_din1}
\end{figure}

So, even if in absolute terms many structural quantities change (the number of nodes, the number of links, the degrees, etc.), under both null models the significance profiles are extremely stable, clearly pointing out that the deviating patterns are systematic and the relative importance of each motif remains constant.

The above results indicate that the ITN is almost completely consistent with a quasi-equilibrium network driven by the local (non-)reciprocated degrees $k_{i}^{\leftrightarrow}$, $k_{i}^{\leftarrow}$ and $k_{i}^{\rightarrow}$.
Even if the latter vary considerably over time, presumably under the effect of complicated economic and political processes (such as the creation of new independent states, globalization and the establishment of reciprocated relationships), once these processes are reabsorbed into the evolution of the local constraints, the quasi-equilibrium character of the network becomes manifest.

\section{The Dutch Interbank Network}

We now turn to the analysis of the DIN. We consider a data set where nodes are Dutch banks and a link from node $i$ to node $j$ indicates that bank $i$ has an exposure larger than 1.5 million euros and with maturity shorter than one year, towards a creditor bank $j$ \cite{mybanks}. We consider 44 quarterly snapshots of the network, from the beginning of 1998 to the end of 2008.
The last year in the sample represents the first year of crisis, i.e. when the recent financial and banking crisis became manifest. 
During the evolution of the DIN, the number of banks and the number of connections (both total and per vertex) changed only moderately \cite{mybanks}. Since the entity of the variation of these quantities is much smaller throughout the evolution of the DIN than in the case of the ITN, we might expect the DIN to display even more stable patterns than the ITN. However, as we now show, the opposite is true.

If we repeat the calculation of the $z$-scores and significance profiles that we used to produce Fig. \ref{fig_wtw}, for the DIN we obtain the corresponding Fig. \ref{fig_din1}. 
What we find is that, unlike the ITN, the DIN displays highly non-stationary profiles, with no collapse of all the different snapshots onto a unique curve. Notice that the moderate change of quantities like the network's size makes the rescaling defining the significance profile practically unnecessary: in fact, the DIN $z$-scores do not differ so much from the DIN significance profiles, as shown in Fig. \ref{fig_din1} \cite{mybanks}. This confirms that the evolution of the triadic profiles is not due to changes in the size of the network, and is a genuine effect.

Many motifs have, in different periods, both positive and negative $z$-scores, indicating a complete inversion of their significance (from under-representation to over-representation and vice versa). 
The large (in absolute value) $z$-scores and their wild temporal fluctuations indicate that, unlike the ITN, the DIN behaves like an out-of-equilibrium network, whose driving dynamics cannot be captured by the selected constraints alone.

\begin{figure}[t!]
\centering
\includegraphics[width=0.388\textwidth]{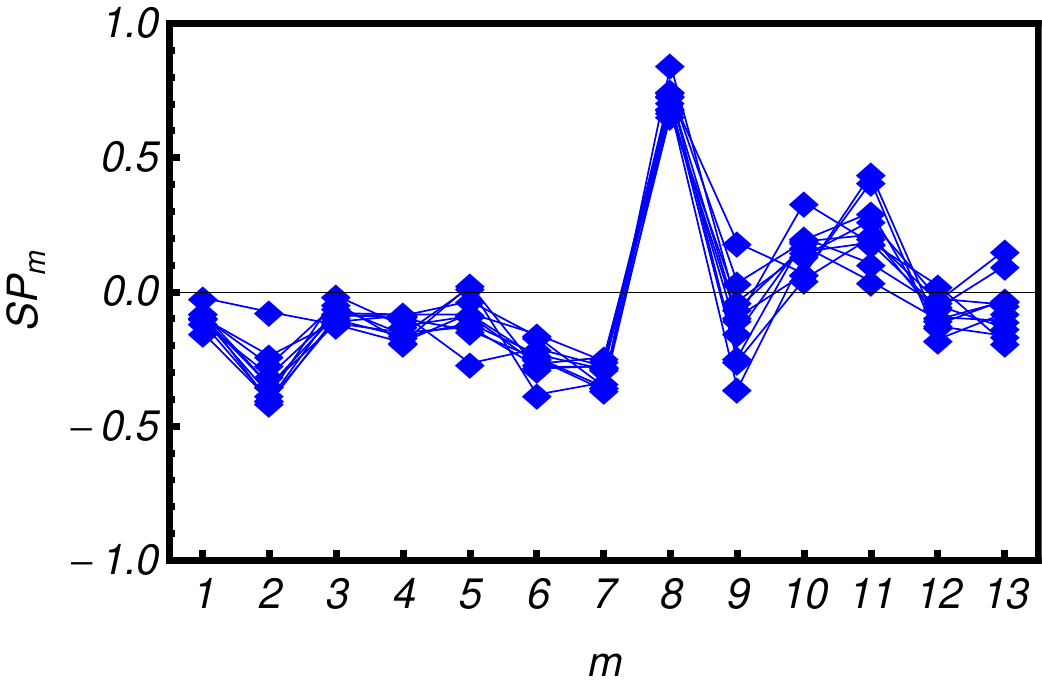}\\
\includegraphics[width=0.388\textwidth]{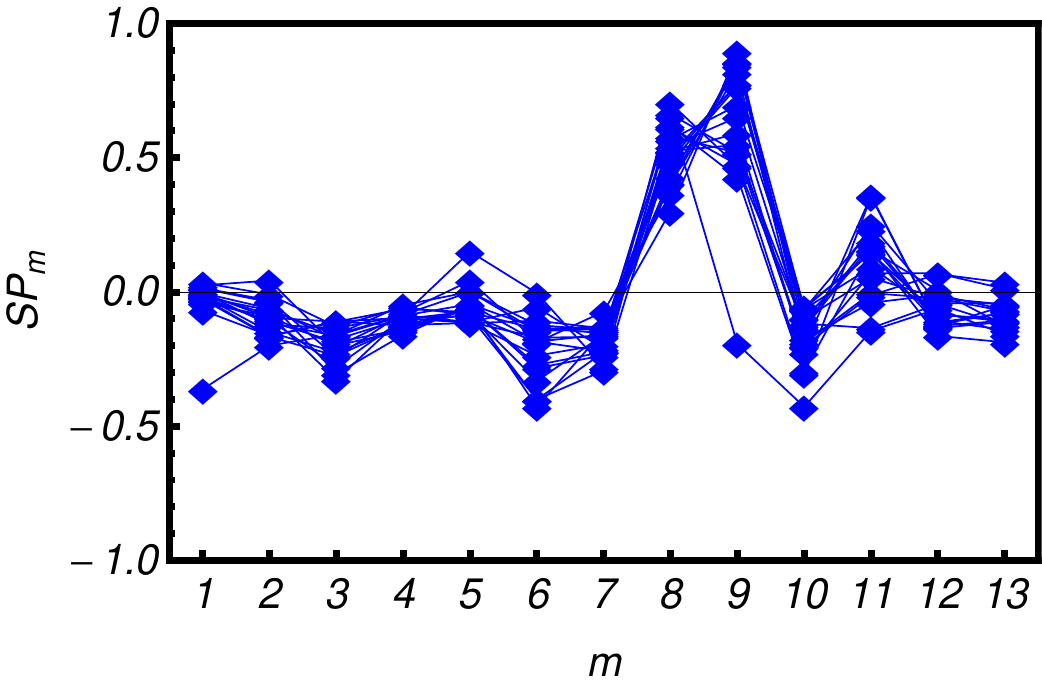}\\
\includegraphics[width=0.388\textwidth]{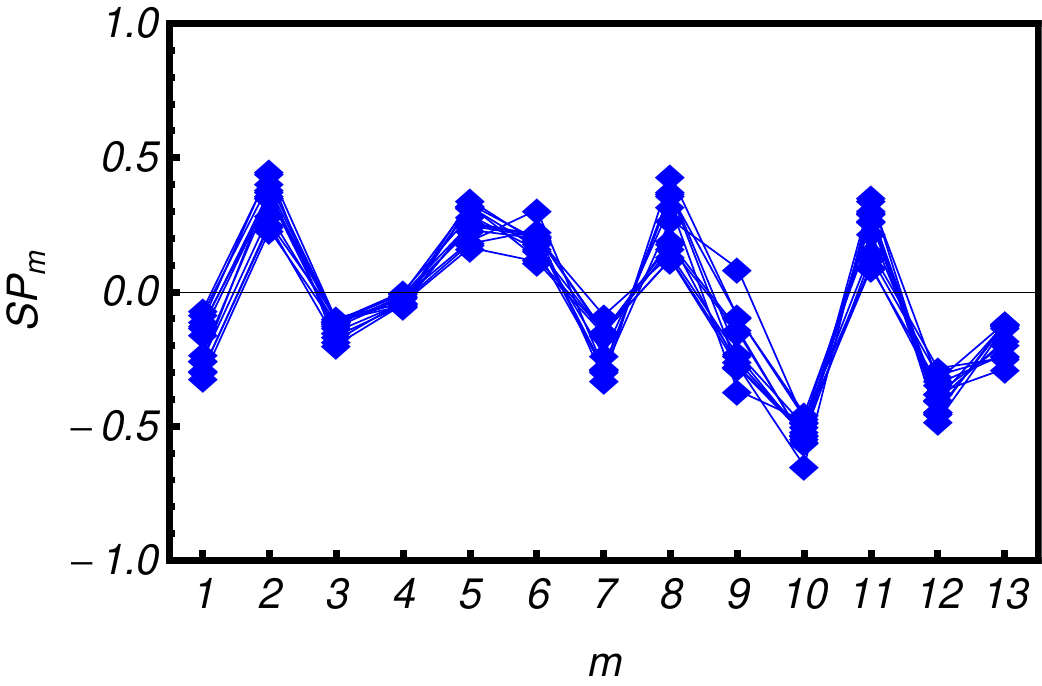}\\
\includegraphics[width=0.388\textwidth]{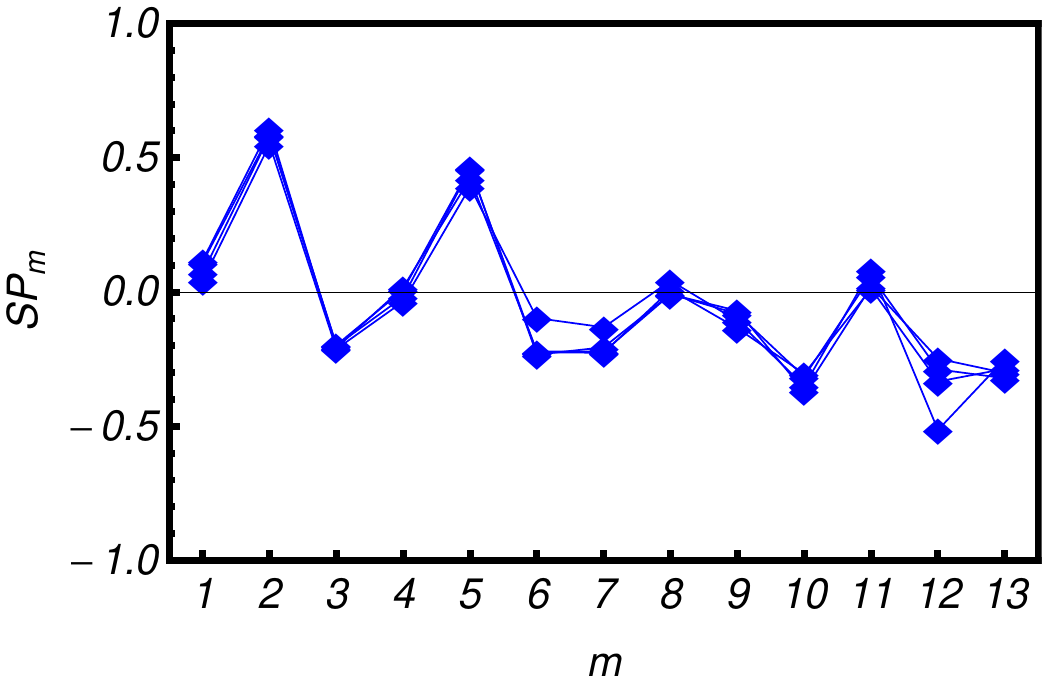}
\caption{Significance profiles of the 13 triadic, binary, directed motifs of the DIN under the DCM ($\textcolor{blue}{\blacklozenge}$) for the four subperiods (from top to bottom) 1998Q1-2000Q2, 2000Q3-2004Q4, 2005Q1-2007Q4 and 2008Q1-2008Q4. \copyright 2014 IEEE. Reprinted, with permission, from Proceedings of the Ninth International Conference on Signal-Image Technology \& Internet-Based Systems (SITIS 2013), pp. 530-537 (edited by IEEE) (2014).}
\label{fig_din2}
\end{figure}

\begin{figure}[t!]
\centering
\includegraphics[width=0.388\textwidth]{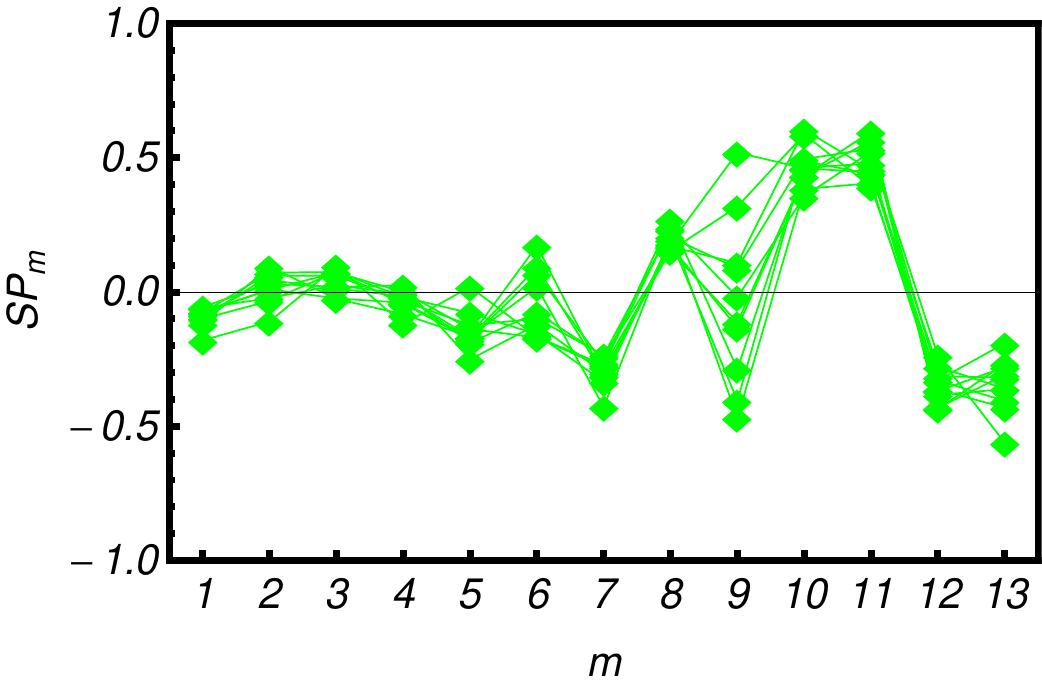}\\
\includegraphics[width=0.388\textwidth]{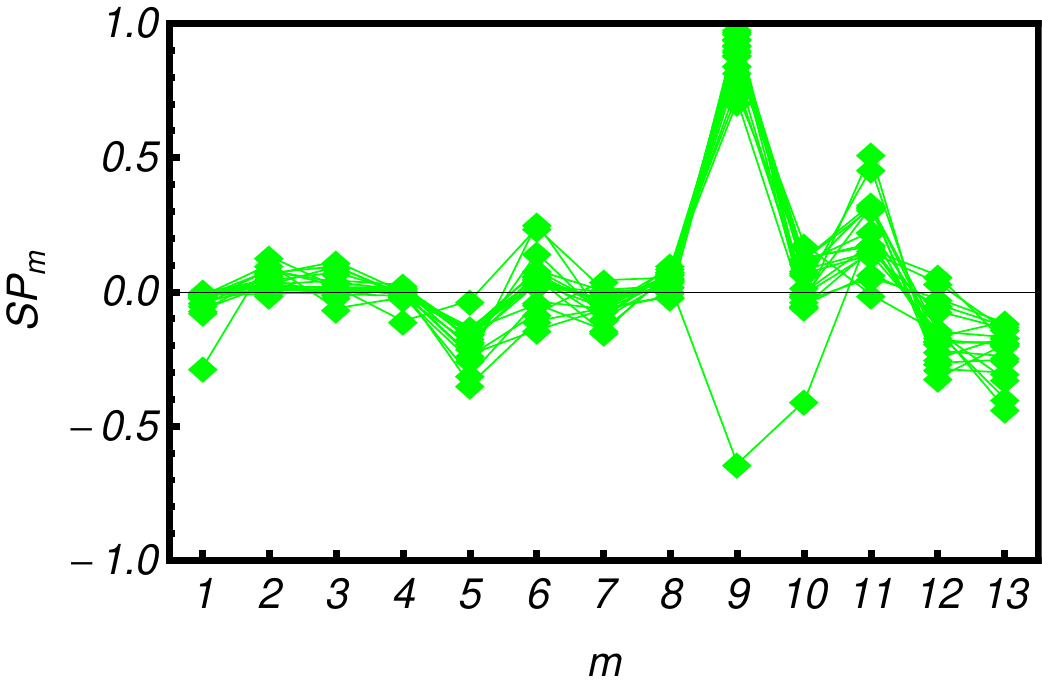}\\
\includegraphics[width=0.388\textwidth]{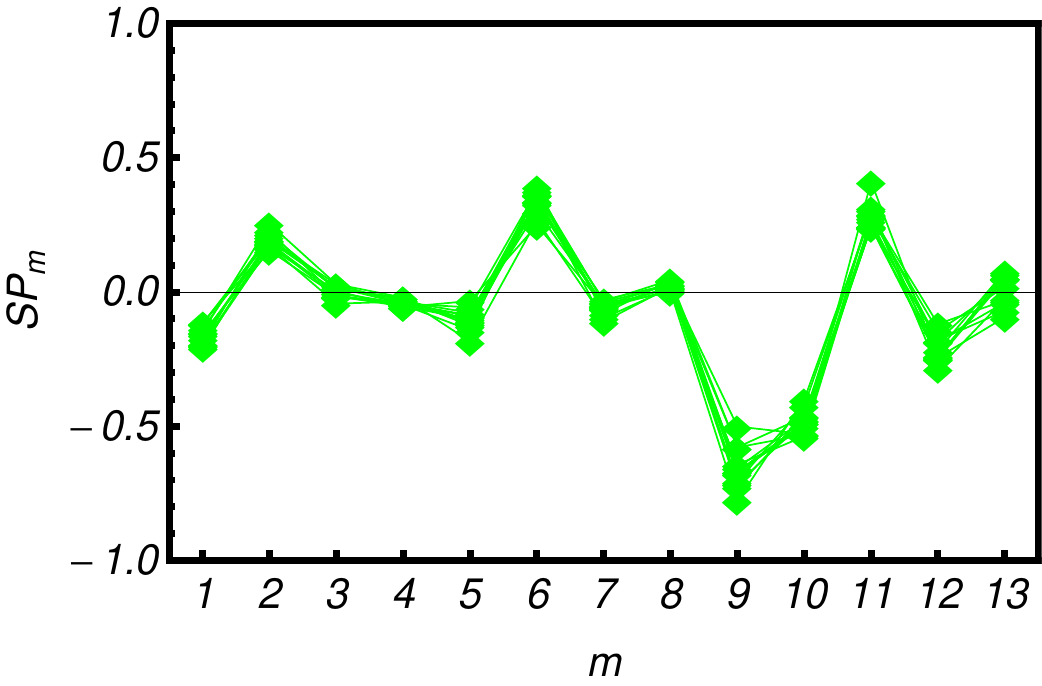}\\
\includegraphics[width=0.388\textwidth]{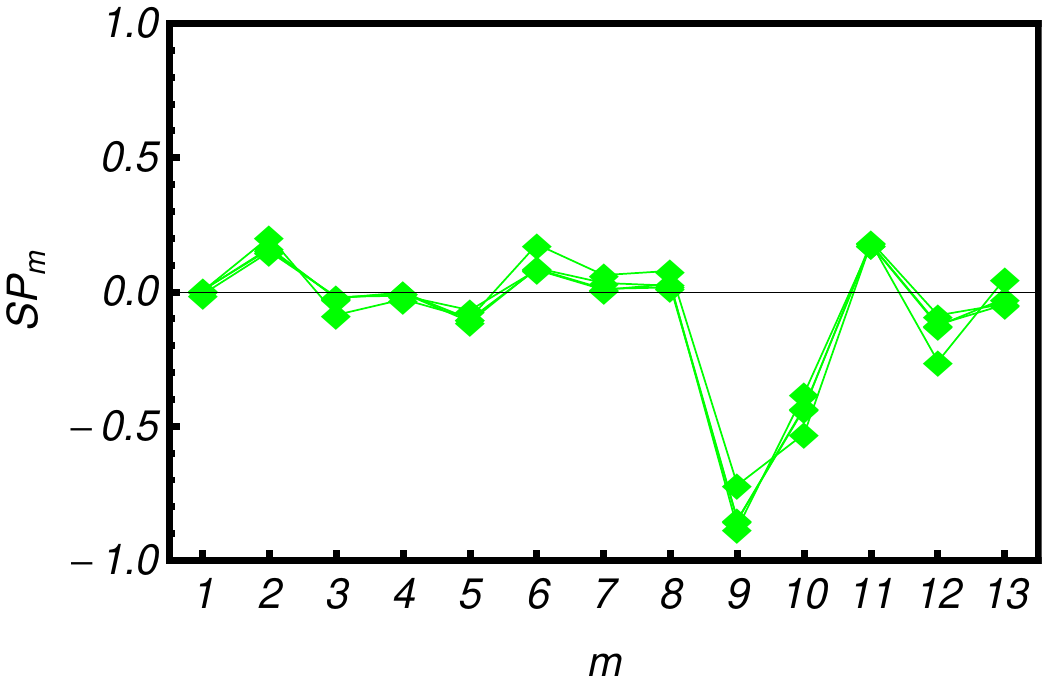}
\caption{Significance profiles of the 13 triadic, binary, directed motifs of the DIN under the RCM ($\textcolor{green}{\blacklozenge}$) for the four subperiods (from top to bottom) 1998Q1-2000Q2, 2000Q3-2004Q4, 2005Q1-2007Q4 and 2008Q1-2008Q4. \copyright 2014 IEEE. Reprinted, with permission, from Proceedings of the Ninth International Conference on Signal-Image Technology \& Internet-Based Systems (SITIS 2013), pp. 530-537 (edited by IEEE) (2014).}
\label{fig_din3}
\end{figure}

However, under both null models we can identify relatively stable triadic profiles if we partition the entire 11-year period into four subperiods.
These periods are 1998Q1-2000Q2, 2000Q3-2004Q4, 2005Q1-2007Q4, and 2008Q1-2008Q4 (where $y$Q$i$ denotes the $i$th quarter of  year $y$). 
This is shown in Fig. \ref{fig_din2} under the DCM and in Fig. \ref{fig_din3} under the RCM. 
Both figures show the four subperiods separately and the almost complete collapse of all snapshots within each subperiod. 

Thus, we can conclude that the overall non-stationary dynamics of the DIN can be approximately decomposed into four relatively stationary phases connected by major structural transitions.
Within each stationary subperiod, considerations analogous to those we made in the example of the ITN may apply.
By contrast, across subperiods major structural changes occur, and the description of the network is irreducible to the change of the bank-specific variables directly affecting the degrees of the corresponding nodes.

\subsection{Early-warning signals}

If we label the fourth subperiod as the `crisis' phase, the triadic profiles of this period can be considered as the `topological fingerprints' of the crisis. It is interesting to notice that these fingerprints were to a large extent anticipated by the significance profiles of the third subperiod (2005-2007).
We might therefore interpret the latter as a sort of latent `pre-crisis' phase.
Remarkably, the most dramatic change of the significance profiles turns out to occur between the second and third subperiods, not between the third (pre-crisis) and fourth (crisis) ones as one might naively expect. 
This indicates that the main structural transition occurred at the beginning of the pre-crisis phase and not at the onset of the crisis itself, suggesting that monitoring the evolution of the triadic profiles could potentially represent a way to detect early-warning signal of interbank crises. 


\begin{figure}[t!]
\centering
\includegraphics[width=0.49\textwidth]{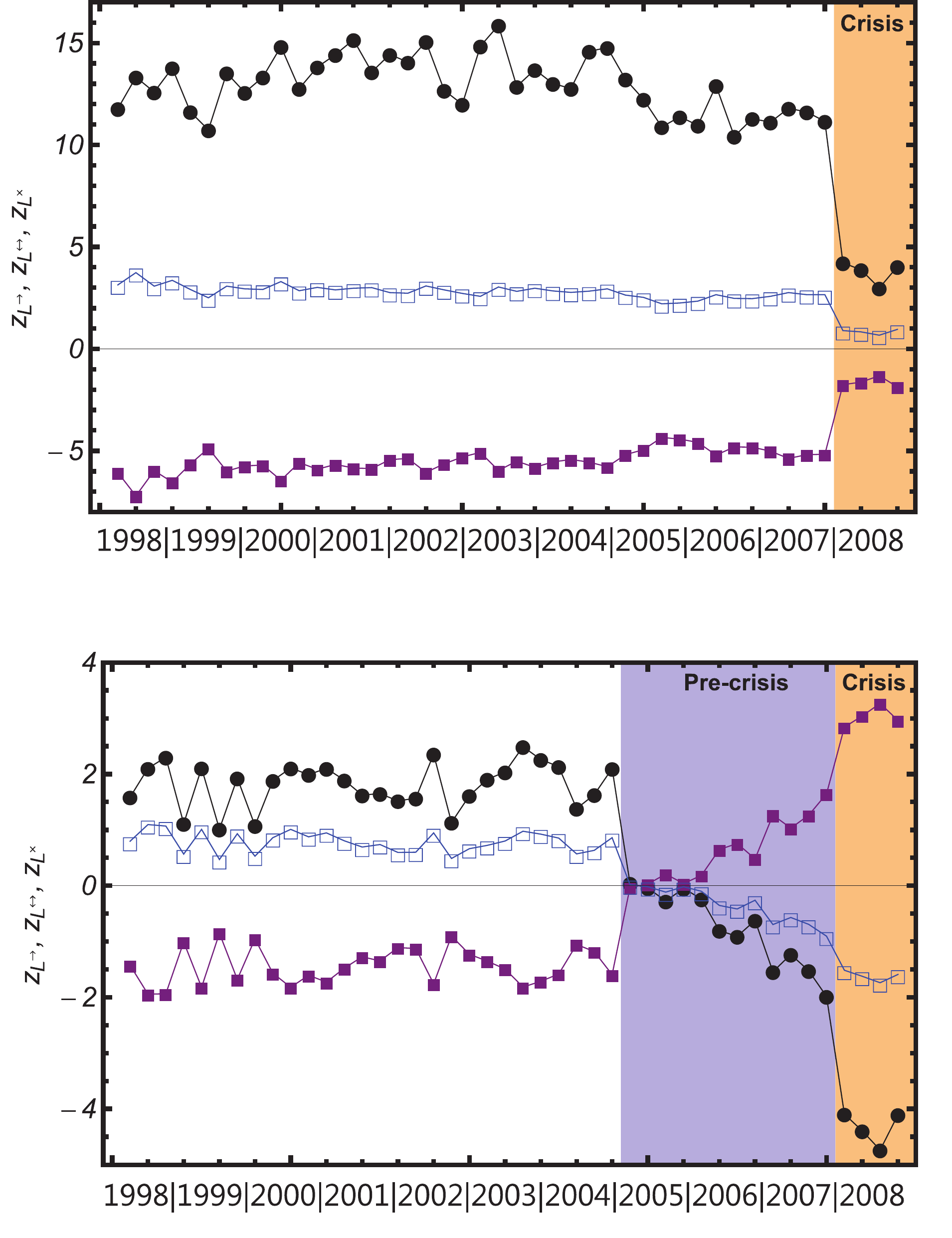}
\caption{$z$-scores of the 3 dyadic, binary, directed motifs ($\textcolor{violet}{\blacksquare}$ - $L^\rightarrow$, $\textcolor{black}{\bullet}$ - $L^\leftrightarrow$, $\textcolor{blue}{\Box}$ - $L^x$) for the 44 quarterly snapshots of the DIN under the DRG (top panel) and the DCM (bottom panel).}
\label{fig_din4}
\end{figure}

The above consideration suggests that an important analysis to perform is plotting the temporal evolution of the $z$-scores over time, for each motif separately.
In this case using the DRG as an additional null model turns out to provide interesting insights.
We recently found that the dyadic $z$-scores of the DIN, if calculated under the DRG, appear to suddenly collapse to their final values only when the crisis is already manifest \cite{mybanks}. 
In particular, the top panel of Fig. \ref{fig_din4} shows the temporal evolution of the 3 dyadic motifs, highlighting the limited role of the homogeneous benchmark in signaling the upcoming event. 
While the DRG correctly identifies the global structural change provoked by the economic crisis (emphasizing that the critical configuration is `anomalous' with respect to the previous decade), it does not provide any useful early-warning signal. 
Note that the fact that the DRG correctly identifies the `crisis' only in terms of dyadic properties is in any case a fundamental result showing that  there are clear signatures of the crisis in the DIN's topology. 
Without this preliminary observation, looking for early-warning signals in the evolution of the dyadic properties themselves would have no empirical justification.

Performing the same analysis under the DCM yields a completely different result \cite{mybanks}. The bottom panel of Fig. \ref{fig_din4} shows that in this case the dyadic $z$-scores undergo a gradual evolution towards the collapsed configuration, thus providing an early-warning signal of the crisis. 
Remarkably, after a period of minor fluctuations, all the trends of the dyadic $z$-scores show a sudden inversion of sign at the beginning of 2005, thus backdating the beginning of the DIN's major structural change three years before its dramatic manifestation in 2008 \cite{mybanks}.

It is very important to check whether the above findings extend to triadic (and in principle higher-order) motifs as well.
Indeed, in over-the-counter (OTC) markets (where transactions between two banks are not disclosed to third parties), triadic motifs are the smallest structural patterns where systemic risk starts to build.
While within a dyad both banks are clearly aware of all the connections existing between them, within a triad each bank is only aware of its connections to and from the other two banks, and not of the possible connections existing among the latter.

\begin{figure}[t!]
\centering
\hspace{1.4cm}\includegraphics[scale=0.6]{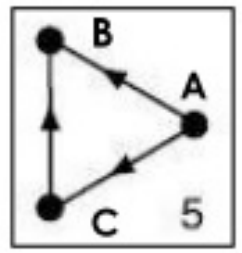}
\includegraphics[width=0.54\textwidth]{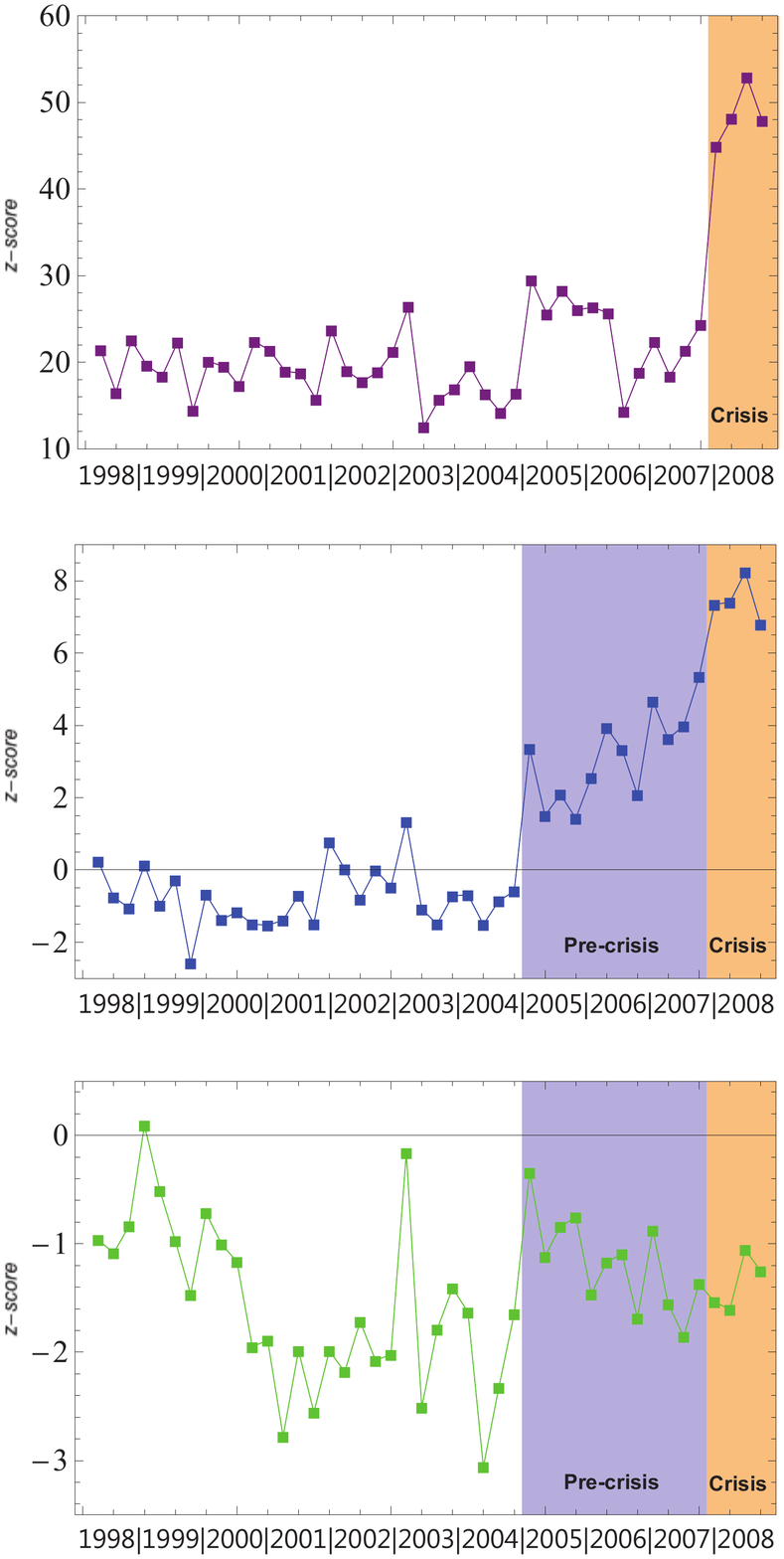}
\caption{$z$-scores of motif $m=5$ under the DRG ($\textcolor{violet}{\blacksquare}$, top panel), the DCM ($\textcolor{blue}{\blacksquare}$, middle panel) and the RCM ($\textcolor{green}{\blacksquare}$, bottom panel).}
\label{fig_din5}
\end{figure}

For instance, as was pointed out e.g. in ref. \cite{co-pierre}, in motif number $m=5$ (see Fig. \ref{fig_din5}) the
bank A is prepared to the direct default of banks B and C, but it is not prepared to the indirect effects of B's default through bank C, precisely because it ignores that B and C are linked.
This can lead to an underestimation of `counterparty risk'.
By looking at  Fig. \ref{fig_din5}, we see that the $z$-score of this particular motif shows very different behaviours under the DCM and the RCM. In the first case, it seems to display the same kind of trend as the dyadic $z$-scores, i.e. to reveal a pre-crisis phase interpretable as an early-warning signal.
However, since any triad is necessarily a combination of three dyads, the reason for the behaviour of a triad as a whole might simply be the combined result of the trends of the three underlying dyads. 
While the DCM is not able to control for this effect, the RCM is, precisely because in this stricter null model the dyadic properties are separately controlled for at each node.
Indeed, Fig. \ref{fig_din5} shows that  the `early-warning' character of motif 5 completely disappears under the RCM, and that the values of the $z$-score are now not very significant.
This proves that this motif is not particularly relevant in itself.

\begin{figure*}[t!]
\centering
\includegraphics[width=0.85\textwidth]{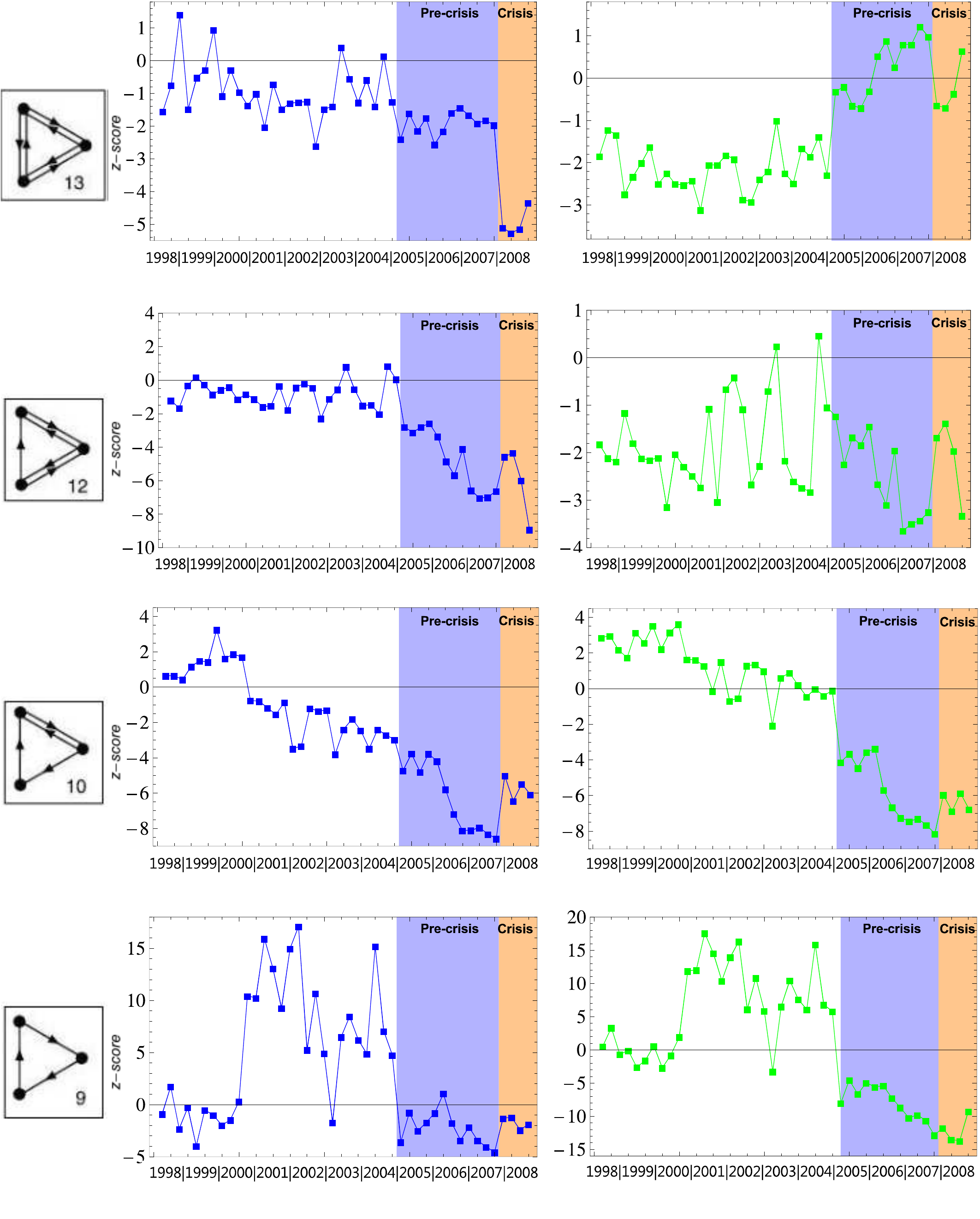}
\caption{$z$-scores of motifs $m=9,\:10,\:12,\:13$ under the DCM ($\textcolor{blue}{\blacksquare}$, left column) and the RCM ($\textcolor{green}{\blacksquare}$, right column).}
\label{fig_din6}
\end{figure*}

The triadic patterns that turn out to be much more relevant to systemic risk are the \emph{directed loops}.
Note that in a circular loop of three banks, each of the three banks involved is not aware of counter-party risk looping back to itself, thus creating additional dependencies among potential defaults not incorporated in their bilateral risk pricing. 
Also note that circularity is not necessarily associated with strong risk externalities by itself, but \textit{unreciprocated circularity} is. For example, within a full dyad risk loops back between the two banks as well; however, both parties are \textit{aware} of it and can properly include the increased correlation in their risk pricing. By constrast, at the triadic level, loops of length three involving an increasing number of reciprocated dyads (i.e. motifs number 9, 10, 12 and 13, see Fig. \ref{fig_din6}) are increasingly less prone to the risk externality. Unreciprocated loops can therefore be considered to be a sort of `autocatalytic risk loops'.
Since longer loops have smaller probabilities of cascading defaults, the most dangerous autocatalytic risk loops are presumably those involving three banks.

In Fig. \ref{fig_din6} we show all the four directed loops, in decreasing order of dyadic reciprocity. 
We note that, while for the less dangerous loops ($m=13$ and $m=12$) the behaviour is similar to that of motif 5, i.e. the $z$-scores are no longer significant under the RCM, the more dangerous loops ($m=10$ and $m=9$) are strongly significant.
These motifs are not just the result of the combination of the participating dyads. Remarkably, the trend of motif $m=9$ leads to the identification of an even `earlier' phase of structural change ranging, approximately, from 2000 to 2005 (i.e. the beginning of the `pre-crisis' phase). This behavior is so peculiar to justify the denomination of this period as `cyclic anomaly phase' \cite{mybanks}. Our previous discussion seems to suggest that, during the `cyclic anomaly' phase, banks might have systematically underestimated risk externalities \cite{mybanks}.

\section{Conclusions}

In the present paper we have proposed a way to investigate whether real economic network are in or out of equilibrium, by introducing the concept of quasi-equilibrium graph ensembles driven by the dynamics of local constraints and relating the stationarity of a network to its \textit{statistical typicality} with respect to a chosen ensemble. So, `stationarity' here does not mean constancy of the numerical values of certain topological quantities across time: it means that the newtork's evolution is systematically driven by the dynamics of the chosen constraints, and so by the process determining the evolution of the constraints themselves. 
So, even if the quantities usually investigated in network theory as the number of nodes and the nodes' degree vary over time, the explanatory power of the chosen constraints may remain constant.

Our empirical results show that the ITN and the DIN display two completely different behaviours: while the ITN is an equilibrium network, the DIN is an out-of-equilibrium one.

The in- and out-degree sequences of the ITN always fail in explaining the triadic structure, although they have been shown to enclose the necessary information to reproduce properties like the degree-degree correlations and the clustering coefficient \cite{pre1}. 
On the other hand, we found that the RCM, which also constrains the numbers of reciprocated links, can replicate the triadic structure almost perfectly. This confirms the important role of reciprocity in economic networks.

It turns out that the 44 temporal snapshots of the DIN do not collapse to a single profile (under both null models) and that four sub-periods with different profiles can be distinguished. The present analysis seems to suggest the following scenario: after the `cyclic anomaly phase', where many risky patterns were established (all the partly reciprocated loops - motifs $m=10,\:12,\:13$ - were much less abundant than the completely unreciprocated loop), banks started to not trust each other anymore. In fact, even if during the `pre-crisis' phase, the loops with small or no reciprocation - motifs $m=10,\:9$ - became increasingly under-represented, also the reciprocated dyads became increasingly under-represented. As explained in \cite{mybanks}, they might have redirected their links, increasing the systemic risk while avoiding mutual interactions and, in so doing, pushing the system towards the critical configuration.

These results seems to indicate that the OTC transactions indeed have the potential to create unintentional, but destabilizing, patterns, feeding into the debate on how OTC markets can be monitored and regulated. This work calls for future studies aimed at understanding the potential of monitoring the non-stationary properties of interbank networks within the framework of bank regulation.

So, even if on one hand the non-stationary character of the DIN  makes the description of the system more complicated than that of equilibrium networks such as the ITN, on the other hand it provides a key piece of information that in this case might open the potential to detect early-warning signals.

\appendix
\section{Directed Random Graph Model}

The \textit{Directed Random Graph Model} (DRG) is the directed version of the Erdos-Renyi random graph \cite{newman_expo}. The only quantity defining the latter is the \textit{total number of links}, $L=\sum_i\sum_{j(\neq i)}a_{ij}$, of a given network. In an economic context, they represent the total number of trading relations observed in the particular system. Given the extreme diversity of the agents playing a role in economic and financial systems, the only (global) constraint defining the DRG cannot be expected to reproduce all the properties of interest. Nevertheless, it can still clearly signal a readjustment of the system's structure of interest taking place at a global level. 

The DRG Hamiltonian is

\begin{equation}
H(\mathbf{A},\vec{\theta})=\theta L
\end{equation} 

\noindent and the resulting probability for the generic network, $\mathbf{A}$, is

\begin{eqnarray}
P(\mathbf{A}|\vec{\theta})&=&\prod_i\prod_{j(\neq i)}p^{a_{ij}}(1-p)^{1-a_{ij}}=\\\nonumber
&=&p^{L}(1-p)^{N(N-1)-L}
\end{eqnarray} 

\noindent where $p\equiv\frac{x}{1+x}$ with $x\equiv e^{-\theta}$ \cite{mymethod}. 
Given a real network $\mathbf{A}^*$, the parameter $x$ can be set to the value $x^*$ that maximize the likelihood of $\mathbf{A}^*$, or equivalently that enforce eq.(\ref{exp}). The latter reads in this case

\begin{eqnarray}
\langle L\rangle=\sum_{i}\sum_{j(\neq i)}\frac{x^*}{1+x^*}=L^*.
\label{drgsys}
\end{eqnarray}

Once the unknown variables are numerically determined, the expected value of any adjacency matrix entry simply becomes $\langle a_{ij}\rangle^*=p^*=\frac{x^*}{1+x^*}$. 
The latter can be used to immediately calculate the expected value $\langle X\rangle^*$ of any topological quantity $X$ of interest \cite{mymethod}. By directly solving eq.(\ref{drgsys}), one finds that the parameter $p$ is nothing else than the network \textit{connectance}, also known as \textit{link density}, i.e. $p^*=\frac{L^*}{N(N-1)}$.

\section{Directed Configuration Model}

In a directed network $\mathbf{A}$, for each node $i$ one can separately define the number $k^{out}_i=\sum_{j(\ne i)}a_{ij}$ of out-going links, or \textit{out-degree}, and the number  $k^{in}_i=\sum_{j(\ne i)}a_{ji}$ of in-going links, or \textit{in-degree}.
The in- and out-degree are the simplest node-specific local properties. They often reflect some nontrivial node-specific dynamics and are typically extremely heterogeneous in real economic networks \cite{preferentiallending}. If the in- and out-degree of all nodes are both included as constraints in the vector $\vec{C}$, one obtains the so-called \textit{Directed Configuration Model} (DCM) \cite{newman_expo}. 

The DCM Hamiltonian is

\begin{equation}
H(\mathbf{A},\vec{\theta})=\sum_{i=1}^N(\alpha_i k_i^{out}+\beta_i k_i^{in})
\end{equation} 

\noindent and the resulting probability coefficient for the generic network, $\mathbf{A}$, simply factorizes as a product over pairs of nodes:
\begin{equation}
P(\mathbf{A}|\vec{\theta})=\prod_{i}\prod_{j(\neq i)}p_{ij}^{a_{ij}}(1-p_{ij})^{1-a_{ij}}
\end{equation} 

\noindent where $p_{ij}\equiv\frac{x_{i}y_{j}}{1+x_{i}y_{j}}$ with $x_i\equiv e^{-\alpha_i}$, $y_i\equiv e^{-\beta_i}$ \cite{mymethod}. 
Given a real network $\mathbf{A}^*$, the parameters $\{x_i\}$ and $\{y_i\}$ can be set to the values $\{x_i^*\}$ and $\{y_i^*\}$ that maximize the likelihood of $\mathbf{A}^*$, or equivalently that enforce eq.(\ref{exp}). The latter reads in this case

\begin{eqnarray}
\left\{ \begin{array}{l}
\langle k_i^{out}\rangle=\sum_{j(\neq i)}\frac{x_i^*y_j^*}{1+x_i^*y_j^*}={k_i^{out}}^*\:\forall\:i\\
\langle k_i^{in}\rangle=\sum_{j(\neq i)}\frac{x_j^*y_i^*}{1+x_j^*y_i^*}={k_i^{in}}^*\:\forall\:i.
       \end{array} \right.
\label{dcmsys}
\end{eqnarray}

Once the unknown variables are numerically determined, the expected value of any adjacency matrix entry simply becomes $\langle a_{ij}\rangle^*=p_{ij}^*=\frac{x_{i}^*y_{j}^*}{1+x_{i}^*y_{j}^*}$. 
The latter can be used to immediately calculate the expected value $\langle X\rangle^*$ of any topological quantity $X$ of interest \cite{mymethod}.
\vspace{5mm}
\section{Reciprocal Configuration Model}

A more stringent choice of local properties in directed networks allows one to distinguish between \textit{reciprocated} and \textit{non-reciprocated} links. For a given node $i$, we might separately count the number $k_{i}^{\rightarrow}$ of non-reciprocated out-going links, the number $k_{i}^{\leftarrow}$ of non-reciprocated in-coming links and the number $k_{i}^{\leftrightarrow}$ of reciprocated (out-going and in-coming at the same time) links. 
Mathematically, these three different `degrees' are defined as $k_{i}^{\rightarrow}\equiv\sum_{j(\neq i)}a_{ij}^{\rightarrow}$,  $k_{i}^{\leftarrow}\equiv\sum_{j(\neq i)}a_{ij}^{\leftarrow}$ and $k_{i}^{\leftrightarrow}\equiv\sum_{j(\neq i)}a_{ij}^{\leftrightarrow}$ respectively, where $a_{ij}^{\rightarrow}\equiv a_{ij}(1-a_{ji})$, $a_{ij}^{\leftarrow}\equiv a_{ji}(1-a_{ij})$ and $a_{ij}^{\leftrightarrow}\equiv a_{ij}a_{ji}$. 

The graph ensemble where each of the above three quantities is specified for every node is known as the \textit{Reciprocal Configuration Model} (RCM)  \cite{mymethod,mygrandcanonical,myreciprocity}. 
Note that, once the three generalized degrees $k_{i}^{\rightarrow}$, $k_{i}^{\leftarrow}$ and $k_{i}^{\leftrightarrow}$ are specified, the `simpler' out- and in-degrees $k_{i}^{out}$ and $k_{i}^{in}$ are automatically specified as well, but the opposite is not true.
In an economic setting, the reciprocity of economic interactions reflects important properties, such as trust or preference.
Separately controlling for reciprocated and non-reciprocated interations means additionally controlling for the heterogeneity of these properties of nodes.

The Hamiltonian defining the RCM is the following:
\begin{eqnarray}
H(\mathbf{A},\vec{\theta})=\sum_{i=1}^N(\alpha_i k_i^\rightarrow+\beta_i k_i^\leftarrow+\gamma_i k_i^\leftrightarrow).
\end{eqnarray}

\noindent Even if the constraints are now non-linear combinations of the adjacency matrix entries,
the probability still factorizes as a product of dyadic probabilities, making the model analitically solvable \cite{mymethod,mymotifs,mygrandcanonical,myreciprocity}.
The maximization of the likelihood function leads to the following system of equations:

\begin{eqnarray}
\left\{ \begin{array}{l}
\langle k_i^{\rightarrow}\rangle=\sum_{j(\neq i)}\frac{x_i^*y_j^*}{1+x_i^*y_j^*+x_j^*y_i^*+z_i^*z_j^*}={k_i^{\rightarrow}}^*\quad\forall\:i\\
\langle k_i^{\leftarrow}\rangle=\sum_{j(\neq i)}\frac{x_j^*y_i^*}{1+x_i^*y_j^*+x_j^*y_i^*+z_i^*z_j^*}={k_i^{\leftarrow}}^*\quad\forall\:i\\
\langle k_i^{\leftrightarrow}\rangle=\sum_{j(\neq i)}\frac{z_i^*z_j^*}{1+x_i^*y_j^*+x_j^*y_i^*+z_i^*z_j^*}={k_i^{\leftrightarrow}}^*\quad\forall\:i
       \end{array} \right.
\label{rcmsys}
\end{eqnarray}
where $x_i\equiv e^{-\alpha_i}$, $y_i\equiv e^{-\beta_i}$, $z_i\equiv e^{-\gamma_i}$.

The addenda in the three equations above correspond to three different probability coefficients, that we denote as $(p_{ij}^{\rightarrow})^*$, $(p_{ij}^{\leftarrow})^*$ and $(p_{ij}^{\leftrightarrow})^*$ respectively.
These coefficients separately specify the probability of having, from node $i$ to node $j$, a non-reciprocated out-going link, i.e. $\langle a_{ij}^{\rightarrow}\rangle^*$, a non-reciprocated in-coming link, i.e. $\langle a_{ij}^{\leftarrow}\rangle^*$, and two reciprocated links, respectively, i.e. $\langle a_{ij}^{\leftrightarrow}\rangle^*$.

\begin{acknowledgments}
D. G. acknowledges support from the Dutch Econophysics Foundation (Stichting Econophysics, Leiden, the Netherlands) with funds from beneficiaries of Duyfken Trading Knowledge BV, Amsterdam, the Netherlands.
\end{acknowledgments}

\end{document}